\documentclass{article}

\usepackage{arxiv}
\usepackage[utf8]{inputenc}
\usepackage[T1]{fontenc}
\usepackage{textcomp}
\usepackage[hidelinks]{hyperref}
\usepackage{url}
\usepackage{booktabs}
\usepackage{amsfonts}
\usepackage{nicefrac}
\usepackage{microtype}
\usepackage[table]{xcolor}
\usepackage{graphicx}
\usepackage{subcaption}
\usepackage{multirow}
\usepackage{makecell}
\usepackage{acronym}
\usepackage{tabularx}
\usepackage{amsmath}
\usepackage{natbib}
\usepackage[capitalize,nameinlink]{cleveref}
\crefname{figure}{Figure}{Figures}
\Crefname{figure}{Figure}{Figures}
\crefname{equation}{Equation}{Equations}
\Crefname{equation}{Equation}{Equations}

\graphicspath{{./images/}}

\acrodef{PD}{Parkinson's Disease}
\acrodef{HC}{Healthy Control}
\acrodef{yHC}{Young Healthy Control}
\acrodef{ICA}{Independent Component Analysis}
\acrodef{FA}{Factor Analysis}
\acrodef{VS}{Version Signal}
\acrodef{IVT}[I-VT]{Identification by Velocity Threshold}
\acrodef{KS}{Kolmogorov-Smirnov}
\acrodef{EM}{Expectation-Maximisation}
\acrodef{VOG}{Video-Oculography}
\acrodef{CERM}{Cyclopean Eye Residual Movement}
\acrodef{SPEM}{Smooth Pursuit Eye Movements}
\acrodef{DTW}{Dynamic Time Warping}

\title{Estimation of the Cyclopean Eye from Binocular Free Visual Explorations}

\author{
 Maria F. Alcala-Durand \\
  Escuela Técnica Superior de Ingenieros de Telecomunicación\\
  Universidad Politécnica de Madrid\\
  \texttt{maria.adurand@upm.es} \\
    \And
 J. Camilo Puerta-Acevedo \\
  Escuela Técnica Superior de Ingenieros de Telecomunicación\\
  Universidad Politécnica de Madrid\\
 \texttt{juancamilo.puerta@upm.es}  \\
  \And
 Julián D. Arias-Londoño \\
  Escuela Técnica Superior de Ingenieros de Telecomunicación\\
  Universidad Politécnica de Madrid\\
  \texttt{julian.arias@upm.es}  \\
  \And
  Elisa Luque-Buzo \\
  Movement Disorders Unit, Neurology Service\\
  Hospital Gregorio Marañón de Madrid\\  \texttt{elisa.luque@salud.madrid.org} \\
  \And
  Francisco J. Grandas-Pérez \\
  Movement Disorders Unit, Neurology Service\\
  Hospital Gregorio Marañón de Madrid\\
\texttt{francisco.grandas@salud.madrid.org} \\
  \And
 Juan I. Godino-Llorente \\
  Escuela Técnica Superior de Ingenieros de Telecomunicación\\
  Universidad Politécnica de Madrid\\
  \texttt{ignacio.godino@upm.es} \\
}

\begin{document}
\maketitle

\begin{abstract}
Binocular video-oculography records separate signals for each eye, yet human visual processing and oculomotor control operate relative to a single, virtual cyclopean eye. This study addresses the estimation of the cyclopean eye during binocular free visual exploration, a target-free paradigm widely used for the screening of neurological disorders. Using eye-tracking data collected at 1 kHz from 113 participants across three cohorts (patients with Parkinson's Disease, age-matched healthy controls, and young controls), we compare two different methods: one based on an Independent Component Analysis and another based on Factor Analysis. The extracted cyclopean signals were evaluated on their ability to preserve clinically critical fixation and saccadic events using both velocity-threshold and pixel-dispersion event detectors, and were validated against references via Dynamic Time Warping. Results suggest that the technique based on Factor Analysis guarantees full identifiability across all subjects and achieves low alignment errors, performing comparably to the traditional Version Signal (point-wise mean) while dynamically adapting to inter-eye asymmetry. Furthermore, the second factor in this technique was empirically null across all fitted models, confirming that a single latent factor fully accounts for the shared binocular gaze dynamics. These findings establish Factor Analysis as an identifiable framework for estimating the cyclopean eye in target-free visual exploration protocols.

\end{abstract}


\section{Introduction}
Binocular \ac{VOG} records two eye-position channels, yet the visual system does not appear to treat them as two independent signals. Under classical accounts, the two eyes act as halves of a single organ, and perceived visual direction is referenced to a single, virtual cyclopean eye situated midway between them rather than to either physical eye~\cite{ono1982cyclopean}. Subsequent psychophysical work has confirmed this account against the rival claim that direction is judged from the viewing eye alone~\cite{ono2002cyclopean,khokhotva2005cyclopean}, and the sighting-dominant eye appears to have no unique functional role in vision~\cite{mapp2003dominant}. The same shared component appears in eye movements: when the two eyes move together in the same direction, the common part of that motion is the version component, as opposed to the vergence component, which is the part where the two eyes move oppositely and reflects their disparity~\cite{ono2002cyclopean,vanleeuwen1999gazeshift}. Treating the two eyes as if they were one is therefore not only a matter of convenience for downstream processing; it reflects how the underlying signal is thought to be generated.

Computationally, the simplest way to build this idea into a signal is to average the two channels point-wise, a construction usually called the \ac{VS}~\cite{hooge2019gaze}. Averaging reduces the variable error of the signal~\cite{hooge2019gaze} and, when participants look at known targets with the head turned or while stepping towards them, gives a gaze error no larger than that of the better single eye~\cite{cui2006gaze}, but it is only appropriate under symmetric binocular behaviour. In a calibration-validation task, in which participants fixate a grid of known targets, a single eye had a smaller systematic error than the average in 25.8\% to 29.5\% of participants, at the cost of a larger variable error~\cite{hooge2019gaze}. Averaging can also add systematic error when binocular alignment is disturbed, as in convergence insufficiency~\cite{hooge2019gaze}, which is also when a shared estimate is most needed. The two channels are not interchangeable in practice either, since the systematic error can differ between the eyes of the same participant~\cite{hooge2019gaze}, and gaze accuracy decreases for targets away from the centre of the display~\cite{pijpaert2025validation}. An earlier work addressed this limitation for a target-driven task, using \ac{ICA} to recover a \ac{CERM} component from binocular gaze recorded during Smooth Pursuit Tests in \ac{PD} patients and \ac{HC}~\cite{luque2024cyclopean}. Because the target-following baseline wander could be removed first, the residual signal was dominated by large, non-Gaussian saccadic events, and this structure made the shared component identifiable. That study explicitly left open whether the same strategy would generalise beyond \ac{SPEM}, to tasks without a target trajectory to exploit.

Free visual exploration, a target-free task in which participants view static images without instructions, has also been studied as a source of oculomotor biomarkers for the automatic screening of \ac{PD}~\cite{alcala2025screening}. That study modelled the two eyes independently throughout, on the grounds that binocular asymmetries could carry disease-relevant information, and identified combining both eyes into a single cyclopean signal as a natural next step. The present work takes up that suggestion directly: it asks whether a cyclopean estimate can be recovered from free-exploration recordings, and if so, whether the resulting signal still captures the fixation and saccade structure that later analyses depend on.

Eye tracking has become a practical instrument for characterising \ac{PD}, and a recent systematic review reports fixation and saccade descriptors, such as saccade velocity and fixation duration, among the measures correlated with disease severity~\cite{culicetto2025review}. This dependence is detailed: classifiers operating on fixation data alone separate patients from controls at the level of the individual~\cite{uribarri2026deep}, and the microsaccadic structure within a single fixation carries diagnostic information~\cite{wang2026microsaccade}. A cyclopean estimate that smoothed this structure away or shifted it between fixations and saccades would be of little use to these analyses, even if it represented the shared component well. Preserving event structure is therefore treated here as a requirement on the estimate rather than a secondary concern. Moreover, binocular coordination is affected by \ac{PD}: in a study of 33 patients, more than half kept their eyes well aligned for less than 80\% of the time under binocular viewing~\cite{gupta2023objective}, so the symmetric behaviour that averaging assumes cannot be taken for granted in this population.

Nevertheless, free exploration removes the structure that made \ac{ICA} effective for \ac{SPEM}. Without a target trajectory to subtract, the residual that carries the shared component under smooth pursuit is unavailable. \ac{ICA} separates the two channels by assuming the underlying sources are non-Gaussian and distinguishable on that basis, and free-exploration recordings do not guarantee that condition~\cite{hyvarinen2000ica}. An alternative to the pointed analysis is \ac{FA}, since it does not rely on a non-Gaussian assumption. It only requires that a shared factor explain a meaningful part of the covariance between the two channels, leaving the channel-specific remainder in a diagonal noise term, so the shared factor is identifiable by construction whether or not the signal is strongly non-Gaussian. This work compares the two strategies directly under free exploration, requiring the estimate to keep the fixation and saccade structure that is clinically relevant in \ac{PD}~\cite{gupta2023objective}, together with a second choice: which fixation and saccade detector to apply to the resulting cyclopean estimate.

The choice of event detector also shapes the results. The algorithms in common use disagree with one another systematically, so the detector partly determines the event statistics that follow from it~\cite{birawo2022review,nir2025differences}. The two dominant families make opposite trade-offs: velocity-threshold detection is more accurate on clean signals, while dispersion-based detection with an adapted threshold retains more of its accuracy as noise increases, mainly by keeping fixation structure stable~\cite{orioma2026adaptive}. Both families, and the thresholds reported for them, were developed on the signal of a single measured eye. A cyclopean estimate is built from the two channels, and the operation that combines them can change the velocity or the spatial spread that a detector relies on, so whether a detector and its threshold carry over to the constructed signal has to be tested rather than assumed.

Beyond the target-driven approach described above, researchers have mostly developed a single cyclopean signal from the two eyes for a different objective than the one pursued here, which is a signal that follows the recording over time and preserves its fixations and saccades. In gaze estimation from face and eye images, where the task is to predict a gaze direction or point of regard, several methods combine the two eyes' image features while weighting them by per-eye reliability, rather than averaging them symmetrically. A convolutional neural network (CNN) with local and global binocular spatial-attention modules has been used to combine left and right eye feature maps for screen gaze-point regression~\cite{cheng2020gaze}, an asymmetric regression network learns which eye is more reliable and relies on it to predict 3D gaze direction~\cite{cheng2020gaze}, and a related network combines multi-head cross-attention over the two eyes with a face-eye interaction module to the same end~\cite{li2025gaze}. All three predict where gaze is directed at a given instant, which is a different problem from recovering a shared gaze signal over time, and all three are trained against gaze points or directions labelled in public datasets, a ground truth that does not exist for free visual exploration. The same asymmetry has been exploited for identification rather than regression, where modelling the two eyes' dynamics jointly, and interpretably, outperforms either eye alone~\cite{zhang2026binocular}. These methods confirm, in a different modality, that treating the two eyes asymmetrically outperforms naive combination, which is consistent with the motivation for \ac{FA} over the \ac{VS} baseline in this work. They are optimised to minimise point-estimation error at a single instant, however, not to preserve the fine fixation and saccade structure, so their fusion strategies do not transfer directly to the present problem.

This work sets out to answer three questions: (1) does the resulting estimate show what it is meant to show? (i.e. the shared binocular component of gaze); (2) does it preserve the characteristics of the signal? (i.e. fixation and saccade structure); and (3) can it be evaluated at all, since no ground truth exists for a quantity that is not physically measurable?

\section{Materials}
Recordings were collected over two years at two hospitals in Madrid, Spain: Hospital Universitario de Fuenlabrada (HUF) and Hospital General Universitario Gregorio Marañón (HGUGM). The population comprises three cohorts: clinically diagnosed \ac{PD} patients previously evaluated by a movement-disorders specialist, age-matched \ac{HC}, and \ac{yHC}. The study enrolled 48 participants with \ac{PD}, one with multiple system atrophy, 48 \ac{HC}, and 21 \ac{yHC}, the last included as a reference for exploration behaviour unaffected by ageing. Five participants were excluded before any analysis: the participant with multiple system atrophy, three whose recordings were marked as not valid, and one with exotropia of the left eye, leaving 46 participants with \ac{PD}, 47 \ac{HC}, and 20 \ac{yHC}. \ac{PD} patients had less than five years of disease evolution and were assessed in the ON medication state.
 
\subsection{Recording and tasks}
\begin{figure}[htbp]
\centering
\includegraphics[width=0.6\linewidth]{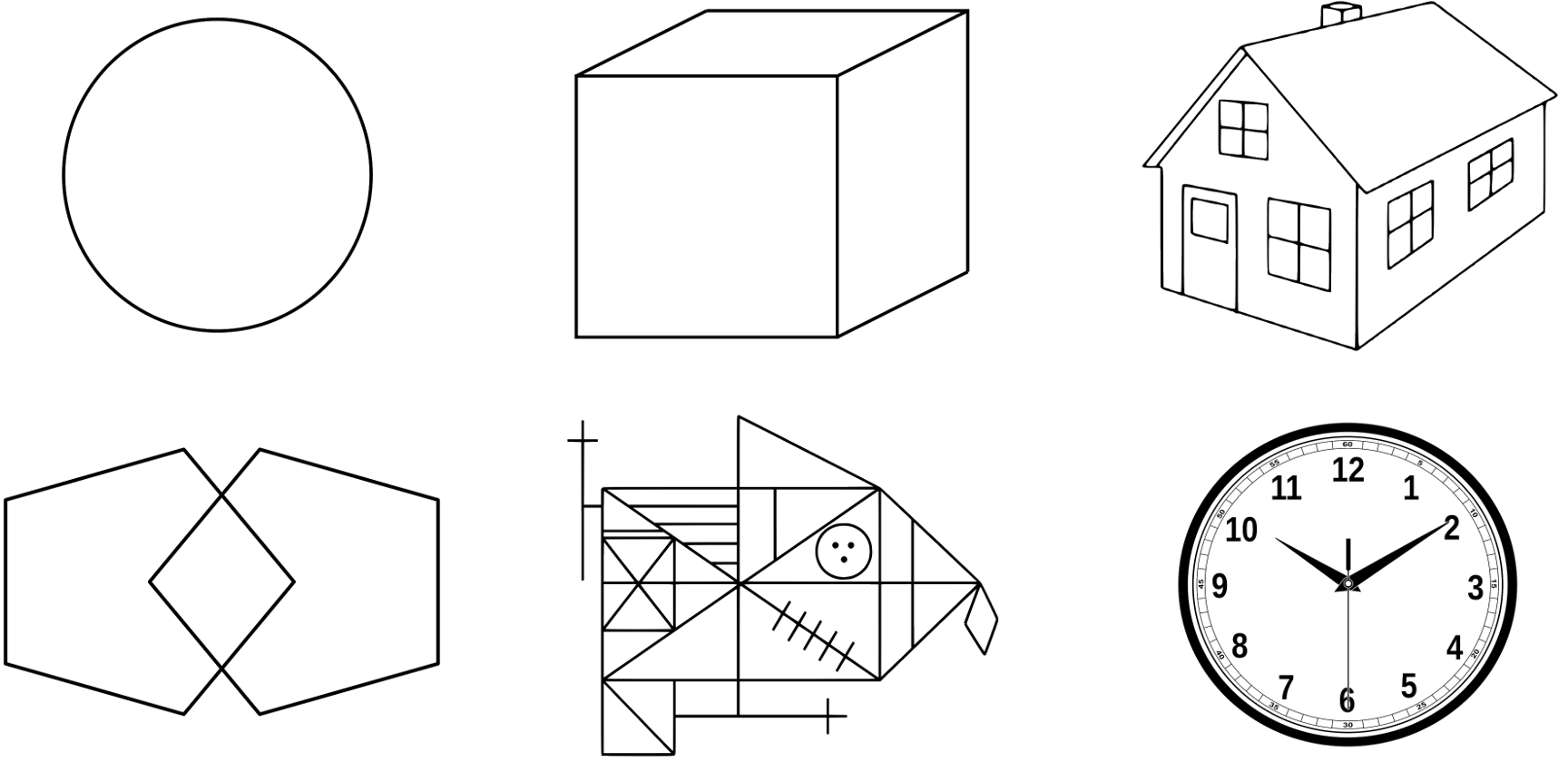}
\caption{The six stimuli, in presentation order (left to right, top to bottom): a circle, a cube, a house, two intersecting pentagons, the Rey-Osterrieth complex figure, and an analogue clock. Each was explored freely for 15~s.}
\label{fig:stimuli}
\end{figure}

Eye movements were recorded with an EyeLink 1000 Plus (SR Research Ltd., Ontario, Canada) at a sampling rate of 1~kHz~\cite{eyelink}. Stimuli were displayed on an LED monitor placed 60~cm in front of the participant. Participants were seated with the head stabilised on a chin rest, and the system was calibrated before each session with a 9-point grid spanning the stimulus area. Each participant explored six images presented sequentially in a fixed order: a circle, a cube, a house, a pair of intersecting pentagons, the Rey-Osterrieth complex figure, and a clock, shown in \cref{fig:stimuli}. Each image was explored for 15~s, and is referred to in the following as Expl.~1 through Expl.~6. Not every recording reaches the analyses below; \cref{sec:preprocessing} sets out which were retained and why.

\subsection{Preprocessing}
\label{sec:preprocessing}
Each sample is labelled by the device parser as belonging to a fixation, a saccade, or a blink, the last detected from pupil occlusion. Blink segments carry unreliable position and velocity and were removed point-wise, discarding on average 6.4\% of the samples of each channel. Gaze coordinates were linearly normalised to the $[0,1]$ range using the device output space, preserving the original aspect ratio. 

Out of the 678 available patient-explorations (spanning six visual explorations, two eyes per participant, and 113 participants), 50 were excluded due to data quality concerns. Each recording was screened against four checks, with the low threshold set at the 1st and the high threshold at the 99th percentile of the retained recordings in every case: (1) insufficient data, where the total measured time falls below the low threshold or blinking events exceed the high threshold; (2) blink related loss, such as a high blink to event ratio, excessive blink duration, or prolonged blink time, all above the high threshold; (3) implausible event counts, where the per eye fixation or saccade count falls below the low or above the high threshold; and (4) inter eye dissimilarity, where the temporal lag of the maximum cross correlation between the left and right channels exceeds the high threshold. A recording was excluded if any criterion was met on either eye, which left 112 participants contributing 628 patient-explorations to the analyses that follow.

\section{Methods}
\label{sec:methods}
 
\subsection{Problem formulation}
\label{sec:formulation}
Let $l_a[n]$ and $r_a[n]$ denote the left and right gaze coordinates along axis $a \in \{x,y\}$ at sample $n$, for a single recording. Assuming, as in the smooth-pursuit case~\cite{luque2024cyclopean}, that both eyes are driven by a common motor command, the two channels can be modelled as a linear mixture of two latent sources,
\begin{equation}
\mathbf{o}_a[n] =
\begin{bmatrix} l_a[n] \\ r_a[n] \end{bmatrix}
= \mathbf{A}_a \, \mathbf{s}_a[n],
\label{eq:mixture}
\end{equation}
where $\mathbf{s}_a[n] = [s_a^1[n], s_a^2[n]]^\top$ holds the two sources: one is the shared point of regard, the cyclopean component, and the other collects eye-specific contributions such as calibration offsets and dominance bias. The two axes are processed independently. Unlike target-driven eye tracking cases, no target trajectory is set, so no baseline wander is estimated or removed and no target-aligned criterion is used to identify the common source.

A naive estimate of the common component is the point-wise mean of the two channels, $\tfrac{1}{2}\big(l_a[n] + r_a[n]\big)$, commonly known as the \ac{VS}~\cite{hooge2019gaze}. Averaging is adequate only under symmetric binocular behaviour, since it assumes that the two eyes move synchronously and with equal gain. If that assumption is broken, averaging can mask or distort the inter-eye differences, for example in saccade dynamics~\cite{hooge2019gaze}; for a substantial minority of individuals a single eye is more accurate than the \ac{VS}~\cite{hooge2019gaze}.

Two methods are evaluated for recovering the common component while assigning the eye-specific discrepancies to a separate term: one based on \ac{ICA}, following \cite{luque2024cyclopean}, whose two sources still need to be distinguished after decomposition, and one based on \ac{FA}, proposed here as an alternative in which the shared component is fixed by construction and requires no such identification step.

\subsubsection{Independent Component Analysis to identify the cyclopean component}
\label{sec:ica}
Following the smooth-pursuit approach, \ac{ICA} inverts \cref{eq:mixture} by seeking two maximally non-Gaussian, statistically independent sources~\cite{hyvarinen2000ica}. The FastICA algorithm was applied per axis with two components, the deflation estimation procedure, the log-cosh contrast, and a fixed random seed; deflation was preferred over the parallel (symmetric) variant because it agreed more often with the expert annotations described below. \ac{ICA} returns the sources without semantic labels and with sign and permutation indeterminacy, so the cyclopean source must be identified after decomposition.

Unlike the smooth-pursuit case, where a moving target isolates a residual with a clear, reliable non-Gaussian signature, free exploration offers no such reference. Identification therefore rests on a single operative criterion: for each candidate source $s_a^c[n]$, $c \in \{1,2\}$, the axis affinity
\begin{equation}
R_a(c) = \tfrac{1}{2}\Big(\rho(s_a^c, l_a) + \rho(s_a^c, r_a)\Big)
\label{eq:affinity}
\end{equation}
was computed, where $\rho$ denotes the Pearson correlation, and the component with the larger $|R_a(c)|$ was taken as the cyclopean source for axis $a$: whichever component more closely resembles what was actually recorded is the most direct evidence available when no target-aligned reference exists to fall back on. Component-wise negentropy under the log-cosh approximation and component variance were also computed, but only as diagnostics reported alongside the automatic choice, not as further selection criteria; \cref{sec:res_ica} shows that they corroborate the correlation-based choice to very different degrees, which is itself informative about why \ac{ICA} struggles under free exploration.

\begin{figure}[htbp]
\centering
\includegraphics[width=\linewidth]{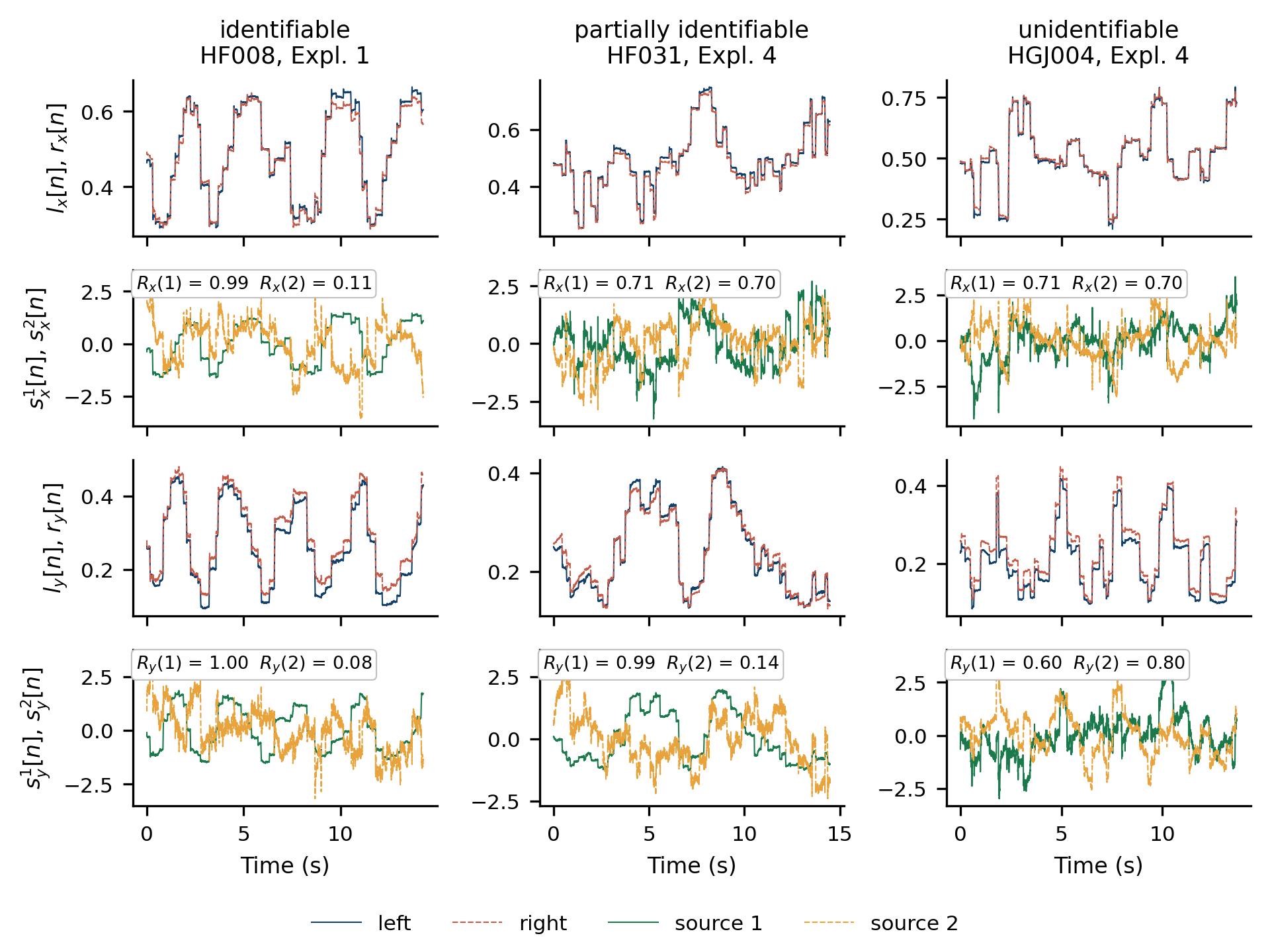}
\caption{Left and right channels and the corresponding \ac{ICA} sources $s_a^1[n]$, $s_a^2[n]$ for three recordings spanning the identifiability spectrum: an identifiable case (patient HF008, Expl.~1, the running example of \cref{sec:res_fa}) where one source tracks the recording and the other is noise-like on both axes; a partially identifiable case (HF031, Expl.~4) where this holds on the vertical axis only; and an unidentifiable case (HGJ004, Expl.~4) where neither axis yields a clearly interpretable pair. Sources are drawn sign-corrected, as they are used in the pipeline, and the affinity $R_a(c)$ of \cref{eq:affinity} is given for both sources on each axis.}
\label{fig:ica_clean_dirty}
\end{figure}

The automatic choice was compared against an expert reference: an annotator inspected the two candidate components for every recording and axis, and labelled as the cyclopean source whichever component visibly reproduced the shape of the two recorded channels, since in the typical case one component tracks the recorded trajectory closely while the other resembles unstructured noise. Recordings were marked as unidentifiable whenever this distinction could not be made, which happened about equally often because both components resembled noise and because both resembled the recording. \Cref{fig:ica_clean_dirty} illustrates this spectrum with three recordings reconstructed from their \ac{ICA} components: one where the two sources separate cleanly on both axes, one where only one axis does, and one where neither axis yields an interpretable pair.

\ac{ICA} recovers each source only up to sign, so the selected component can emerge inverted; therefore, it was sign-corrected so that it correlates positively with the recorded channels, then rescaled by a min-max map, as in the smooth-pursuit work~\cite{luque2024cyclopean}, here onto the amplitude range $[\min(l_a,r_a),\,\max(l_a,r_a)]$ of the recording, since \ac{ICA} recovers each source only up to an arbitrary scale. Loading symmetry was quantified from the row of the unmixing matrix corresponding to the selected component, that is, from the weights with which each eye enters the selected source. Writing $w_l$ and $w_r$ for the left and right weights, the asymmetry index is
\begin{equation}
\alpha = \frac{\big| |w_l| - |w_r| \big|}{|w_l|+|w_r|},
\label{eq:asym}
\end{equation}
which is $0$ for a perfectly balanced binocular component and approaches $1$ when one eye dominates.

\subsubsection{Factor Analysis to identify the cyclopean component}
\label{sec:fa}
\ac{FA} models the observed channels as a small number of shared latent factors plus per-variable unique noise~\cite{barber2012bayesian},
\begin{equation}
\mathbf{o}_a[n] = \boldsymbol{\Lambda}_a \, \mathbf{f}_a[n] + \boldsymbol{\mu}_a + \boldsymbol{\varepsilon}_a[n],
\qquad \operatorname{Cov}(\boldsymbol{\varepsilon}_a) = \boldsymbol{\Psi}_a,
\label{eq:fa}
\end{equation}
with $\boldsymbol{\Psi}_a$ diagonal, fitted by maximum likelihood~\cite{barber2012bayesian}. What makes this model well suited to the cyclopean problem follows directly from how it is built. The factor vector $\mathbf{f}_a[n] = [f_a^1[n], f_a^2[n]]^\top$ is fitted to explain only the part of the signal that $l_a[n]$ and $r_a[n]$ have in common; anything specific to a single eye, including its own measurement noise, is left out of $\mathbf{f}_a[n]$ and absorbed by $\boldsymbol{\Psi}_a$ instead. This matches the cyclopean hypothesis directly, since the quantity of interest is the part of the signal the two eyes share, not the direction of greatest overall spread, which could equally be driven by a single eye moving with unusually large amplitude.

As with \ac{ICA}, different estimation procedures for \ac{FA} carry different assumptions, and the maximum-likelihood approach used here is the most common. Its large-sample theory assumes approximate normality of the observed variables. Under non-normality the maximum-likelihood parameter estimates remain consistent but lose efficiency, and the model $\chi^2$ statistic becomes increasingly inflated; a Monte Carlo study of these effects took univariate skewness of $2.0$ and kurtosis of $7.0$ to represent moderate non-normality~\cite{curran1996robustness}, and these values are used below as reference bounds. Departures from normality are also less of a concern here than in the small samples that motivate most such guidance: departures from normality are reported to have a substantial impact in samples of $50$ or fewer observations and a negligible one from $200$ onwards~\cite[p.~95]{hair2019multivariate}, and each recording contributes on the order of $10^4$ samples per axis, far above either figure.

\begin{figure}[htbp]
\centering
\includegraphics[width=\linewidth]{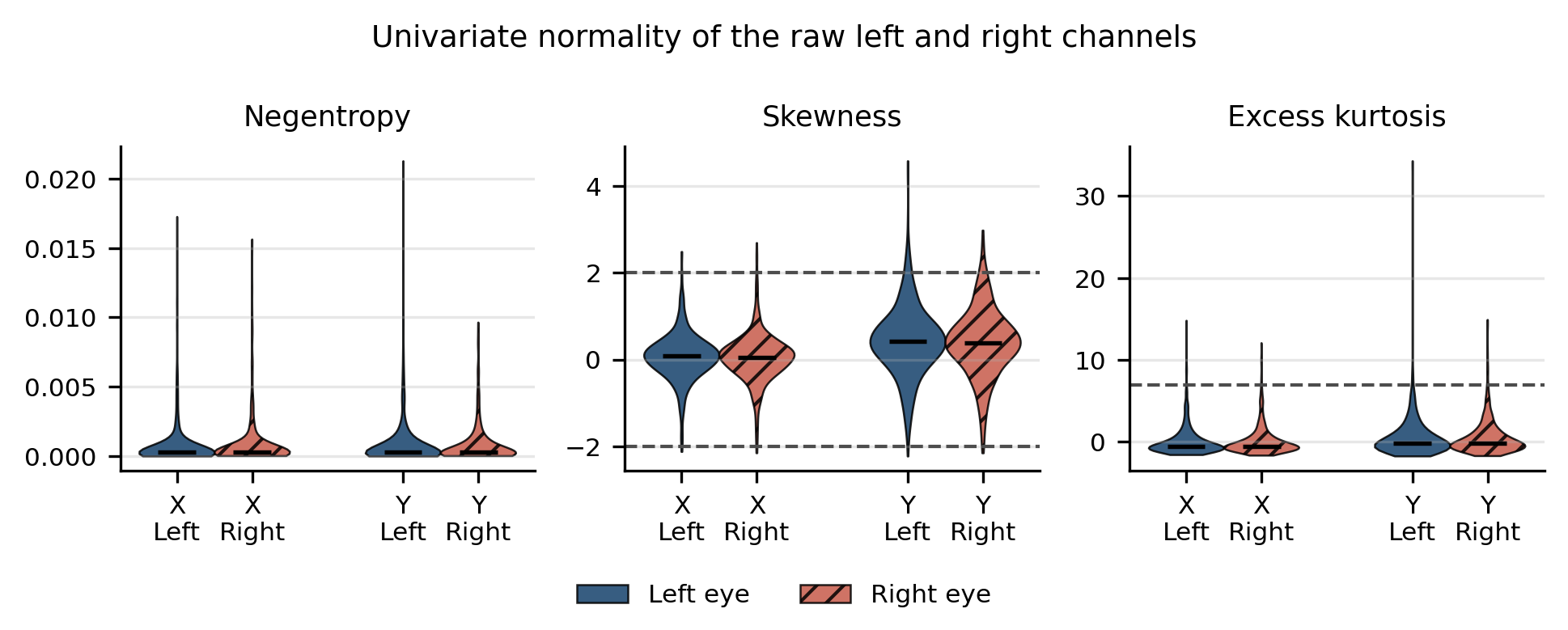}
\caption{Negentropy, skewness, and excess kurtosis of the raw left- and right-eye channels, by axis, aggregated over the corpus. Dashed lines mark the skewness ($|{\cdot}|=2.0$) and kurtosis ($7.0$) values taken to represent moderate non-normality~\cite{curran1996robustness}.}
\label{fig:gaussianity_check}
\end{figure}

The normality of the observed channels was verified empirically. Negentropy, skewness, and excess kurtosis (Gaussian $=0$ for both) were computed for the raw left- and right-eye channels, separately per axis. \Cref{fig:gaussianity_check} shows the resulting distributions: skewness and kurtosis concentrate near zero on both axes, and only $1.4\%$ ($x$) and $3.2\%$ ($y$) of channels exceed the skewness or kurtosis bound above. \ac{FA}'s normality assumption is therefore well supported by the data it is applied to.

\ac{FA} was fitted per axis with two factors and a fixed seed. The first factor is, by construction, the shared binocular component and is taken as the cyclopean estimate without any labelling heuristic; the second factor is retained to test whether a single shared factor is sufficient to describe the recording.

Like an \ac{ICA} component, a factor score has no inherent sign, since negating a factor together with its loadings leaves \cref{eq:fa} unchanged. Rather than correct this after the fact, the cyclopean estimate is built directly from the first-factor loadings and the original channels. Assuming $\boldsymbol{\Lambda}_{a}=[\boldsymbol{\lambda}_a^1,\boldsymbol{\lambda}_a^2]$, where $\boldsymbol{\lambda}_a^1=[\Lambda_{a,l}^1,\Lambda_{a,r}^1]^\top$ is the vector containing the first-factor loadings on the left and right channels, respectively, the cyclopean component is defined as
\begin{equation}
\hat{c}_a[n] = w_l\, l_a[n] + w_r\, r_a[n],
\qquad w_l = \frac{|\Lambda_{a,l}^1|}{|\Lambda_{a,l}^1|+|\Lambda_{a,r}^1|},\ w_r = \frac{|\Lambda_{a,r}^1|}{|\Lambda_{a,l}^1|+|\Lambda_{a,r}^1|},
\label{eq:facyc}
\end{equation}
This differs from the \ac{ICA} case in how amplitude is handled. An \ac{ICA} source has an inherent amplitude ambiguity, since it is recovered only up to an arbitrary scale, so its amplitude carries no physical meaning. Since $\hat{c}_a[n]$ combines $l_a[n]$ and $r_a[n]$ with weights that sum to one, it keeps the physical meaning of the amplitude of the recorded channels, in their native units. The two loadings of the first factor share the same sign in all fitted models, so taking their absolute values only removes the global sign ambiguity of the factor. The same construction does not carry over to \ac{ICA}, whose cyclopean estimate is the selected source itself: the two unmixing weights of that source have opposite signs in 83.0\% of recordings and axes, so taking their absolute values would change the source rather than only its sign. Its two mixing coefficients do share a sign in every recording, but weighting the two eyes by them gives a near-balanced average (median asymmetry $0.016$ on $x$ and $0.028$ on $y$), close to the \ac{VS}, rather than the \ac{ICA} source.

The cyclopean estimate is a weighted combination of the two eyes with weights $w_l, w_r$ fixed by how much each channel contributes to the shared factor, reducing to the \ac{VS} when the two eyes load equally and departing from it, in either direction, only as far as the recording's own binocular balance warrants. Its loading asymmetry is measured with the index $\alpha$ of \cref{eq:asym}, applied to $w_l$ and $w_r$.

\subsection{Detection of fixations and saccades}
\label{sec:events}
A cyclopean estimate, regardless of the method used to obtain it, is useful only if it preserves the oculomotor structure of the recording, in particular the fixation and saccadic pattern on which the characterisation of visual exploration in \ac{PD} depends~\cite{culicetto2025review}. The cyclopean eye is not something the tracker measures; it is a virtual channel built from the two recorded eyes, so there is no true cyclopean fixation or saccade to recover, only the per eye events the estimate should resemble. We therefore assessed each estimate by detecting fixations and saccades on it and comparing the resulting fixation sequences against those of the device parser, applied independently to the left and right channels. These events are not a ground truth, since none exists for a constructed channel, but a common reference, the same for every estimate and detector. The device parser runs on the tracker during recording and cannot be applied to a signal constructed afterwards, so the cyclopean estimates are parsed with separate detectors. As a baseline, the device events of the two eyes are also compared with each other. How a fixation or a saccade is defined matters here, because a rule tuned to a single measured eye need not carry over to a constructed one: the operation that combines the two channels can change the quantity the rule keys on. For this reason, we apply two detectors that key on different quantities: one on gaze velocity (\ac{IVT}) and one on the pixel area a candidate fixation occupies, since fixation detection is a modelling choice rather than a fixed ground truth~\cite{birawo2022review}. The aim is to keep the cyclopean events as close as possible to the per-eye events under each definition and to see which definition carries over to the constructed channel. Thresholds reported in the literature for either detector were tuned on raw per-eye signals, not on a derived cyclopean estimate, so there is no reason to expect them to transfer unchanged, and the operating point for each detector was re-selected on the cyclopean signals themselves.

For each recording, the fidelity of the detected fixations was measured with \ac{DTW}~\cite{sakoe1978dynamic}, which compares the fixation sequences themselves rather than summary statistics such as their number or mean duration. On the samples where both eyes are valid, each signal was coded as a binary sequence $b[n]$, equal to $1$ inside a fixation and $0$ otherwise: $b_{\mathrm{cyc}}[n]$ for the cyclopean estimate, and $b_l[n]$ and $b_r[n]$ for the left and right reference. Two sequences $p[n]$ and $q[n]$ of length $N$ were aligned through the cumulative cost
\begin{equation}
D[i,j] = |p[i] - q[j]| + \min\big(D[i{-}1,j{-}1],\ D[i{-}1,j],\ D[i,j{-}1]\big), \qquad |i - j| \le \tau,
\label{eq:dtw}
\end{equation}
where $i$ and $j$ are the sample indices of $p[n]$ and $q[n]$, so that $\mathrm{DTW}(p[n],q[n]) = D[N{-}1,N{-}1]$ counts the samples whose labels still disagree after the best monotone alignment. With $\tau = 100$ samples (corresponding to 100~ms at 1~kHz), a fixation may start or end up to 100~ms away from its counterpart at no cost, but cannot be matched to a fixation at a different time. The distances to the two device channels were averaged,
\begin{equation}
e_{\mathrm{DTW}} = \frac{\mathrm{DTW}(b_{\mathrm{cyc}}[n], b_l[n]) + \mathrm{DTW}(b_{\mathrm{cyc}}[n], b_r[n])}{2N},
\label{eq:dtwerr}
\end{equation}
and reported as a percentage of the samples of the recording. The inter-eye baseline, $e_{\mathrm{DTW}}^{lr} = \mathrm{DTW}(b_l[n], b_r[n])/N$, is computed in the same way between the two device sequences. For each detector, the threshold minimising the mean of $e_{\mathrm{DTW}}$ over the recordings was selected independently for the \ac{ICA} and \ac{FA} estimates and for the \ac{VS}, included as a baseline that needs no decomposition, so that each method is compared at its own best operating point rather than at a single shared threshold.

\subsubsection{Velocity-threshold detection}
\label{sec:ivt}
The \ac{IVT} algorithm~\cite{salvucci2000ivt} classifies each sample from the instantaneous gaze speed,
\begin{equation}
v[n] = \frac{\sqrt{\big(g_x[n{+}1]-g_x[n]\big)^2 + \big(g_y[n{+}1]-g_y[n]\big)^2}}{\Delta t[n]/1000},
\label{eq:speed}
\end{equation}
in normalised units per s, where $g_x[n]$ and $g_y[n]$ are the coordinates of the signal under analysis, $t[n]$ is the time stamp of sample $n$ in ms, and $\Delta t[n] = t[n{+}1]-t[n]$. Each sample $n$ is labelled a saccade if $v[n] \geq v_\theta$ and a fixation otherwise; consecutive same-class samples were merged into events, and events shorter than 80~ms (fixations) or 6~ms (saccades) were discarded. The threshold $v_\theta$ was swept from 5 to 75~\textdegree/s in steps of 5~\textdegree/s, expressed in normalised units under an assumed horizontal extent of 50\textdegree. \ac{IVT} is velocity-only, whereas the device parser combines velocity with acceleration, so the device's own threshold is not expected to transfer directly to a derived signal; sweeping $v_\theta$ rather than reusing it is a direct test of that.

\subsubsection{Pixel-dispersion detection}
\label{sec:pixel}
As an alternative to a velocity-threshold decision criterion, a dispersion-based detector adapted from \cite{buscher2008eye} was also applied. While \ac{IVT} classifies each interval by speed, this detector classifies by spatial spread: a fixation groups consecutive samples that remain within a bounded region, and the intervals between fixations are labelled saccades. Detection proceeds in two stages. A fixation is initiated once a run of consecutive samples spanning $80$-$100$~ms all fall within a square initiation window of side $d$ pixels. It is then extended sample by sample while each new sample keeps the accumulated set within a larger square window of side $1.5d$ pixels, re-centred on the fixation, with brief excursions outside this window tolerated as noise before the fixation is closed. As in \ac{IVT}, consecutive samples were merged into events and fixations shorter than $80$~ms were discarded. The original algorithm was defined for 50~Hz reading data, so the minimum-sample count was rescaled by the ratio of sampling rates to preserve the same temporal window at the 1~kHz sampling rate used here. The initiation size $d$ was swept from 5 to 50~px in steps of 5~px, the dispersion analogue of the velocity sweep above.

\section{Results}
\label{sec:results}

This section evaluates the identifiability of the \ac{ICA} common component, the properties of the \ac{FA} estimate, and the preservation of fixation structure through event-level validation. The results show that \ac{ICA} frequently yields no clearly identifiable cyclopean component under free exploration, whereas \ac{FA} provides an identifiable shared component across all recordings. At the event level, \ac{FA} and the \ac{VS} achieve comparable \ac{DTW} errors, substantially lower than those of \ac{ICA}, although both produce longer fixation durations than the device parser.

\subsection{Identifiability of the \ac{ICA} common component}
\label{sec:res_ica}
Under free exploration, the \ac{ICA} sources were frequently impossible to label. The expert annotator could not assign the cyclopean component in 13.3\% of recordings on the horizontal axis and 19.8\% on the vertical axis, meaning that 30.2\% of the recordings were ambiguous on at least one axis. This is a genuine identifiability gap rather than a shortcoming of the automatic rule: restricted to the recordings the annotator could confidently label, the correlation-based automatic selection of \cref{eq:affinity} agreed with the annotator in 98.5\% of cases on $x$, 97.4\% on $y$, and 98.0\% pooled across axes. Thus, the two sources of failure are separate, and it is the fraction of recordings with no confident answer at all, rather than a poorly performing automatic rule, that limits \ac{ICA} here. This contrasts with the smooth-pursuit setting, where the best-performing automatic index agreed with the expert labels in 99.5\% of cases~\cite{luque2024cyclopean}. That protocol, however, did not cover the axes orthogonal to the target motion, where both sources also appeared to be noise.

The corroborating diagnostics computed alongside the automatic selection, negentropy, and component variance under the log-cosh approximation do not resolve the ambiguous recordings and, more notably, do not consistently agree with the correlation-based choice even on the confidently labeled ones: the higher-variance component coincides with the chosen one in only 40.0\% of cases on $x$ and 36.5\% on $y$, which is worse than chance, whereas a \ac{KS} distance from Gaussianity agrees much more often, in 84.1\% ($x$) and 80.4\% ($y$) of cases. This is a reversal of the smooth-pursuit setting, where variance was the single most reliable index~\cite{luque2024cyclopean}. Across recordings, the variance of the selected and rejected components is almost exactly the same, so the index carries no information about which is which, whereas the \ac{KS} distance separates them.

The loading asymmetry of the selected \ac{ICA} component (\cref{eq:asym}) has a median of $0.178$ ($x$) and $0.243$ ($y$), and $0.851$ and $0.877$ at the 95th percentile. It exceeds $\alpha = 0.5$, a loading split more extreme than 75/25, in 19.6\% of recordings on $x$ and 23.9\% on $y$.

\subsection{The Factor Analysis estimate}
\label{sec:res_fa}
\ac{FA} removes the identification problem. By construction, the first factor is the shared binocular component, so no recording is unidentifiable. The same expert protocol used for \ac{ICA} was applied here: the annotator inspected every recording, and in every case, the first-factor component satisfied the identification rule, visibly reproducing the recorded channels rather than resembling noise. Agreement is therefore complete not only because the second factor is null, but because the shared component meets the same criterion required of an \ac{ICA} source. Across the corpus, the second-factor loadings were exactly zero in every decomposition on both axes, with no non-zero entries among the $628 \times 2$ fitted models. The second factor is therefore empirically null, which indicates that a single shared factor is sufficient to describe the common binocular motion and that the per-eye discrepancies are captured by the unique-noise term $\boldsymbol{\Psi}_a$ rather than by a second factor.

With the second factor dropped, \cref{eq:fa} reduces to
\begin{equation}
l_a[n] = \Lambda_{a,l}^1\,f_a^1[n] + \mu_{a,l} + \varepsilon_{a,l}[n], \qquad
r_a[n] = \Lambda_{a,r}^1\,f_a^1[n] + \mu_{a,r} + \varepsilon_{a,r}[n],
\label{eq:fareduced}
\end{equation}
so each eye is the same shared factor plus its own residual, $\varepsilon_{a,l}[n]$ or $\varepsilon_{a,r}[n]$, uncorrelated with the other eye's. This residual, not a second shared factor, is where anything specific to one eye is absorbed, whether calibration offset, tracking noise, or a genuine small difference in how the two eyes moved. The null second factor, therefore, means that, in these recordings, whatever the two eyes disagree on looks like independent per-eye noise, not like a second, coherent, shared source of binocular motion. The estimated noise variance of each channel equals the fraction $1 - \rho_a$ of its variance that the other eye does not share, where $\rho_a$ is the correlation between the two eyes on axis $a$; across the corpus it is a median of $0.40\%$ ($x$) and $0.75\%$ ($y$) of each channel's variance, and $3.5\%$ and $11.3\%$ at the 95th percentile.

\begin{figure}[htbp]
\centering
\includegraphics[width=\linewidth]{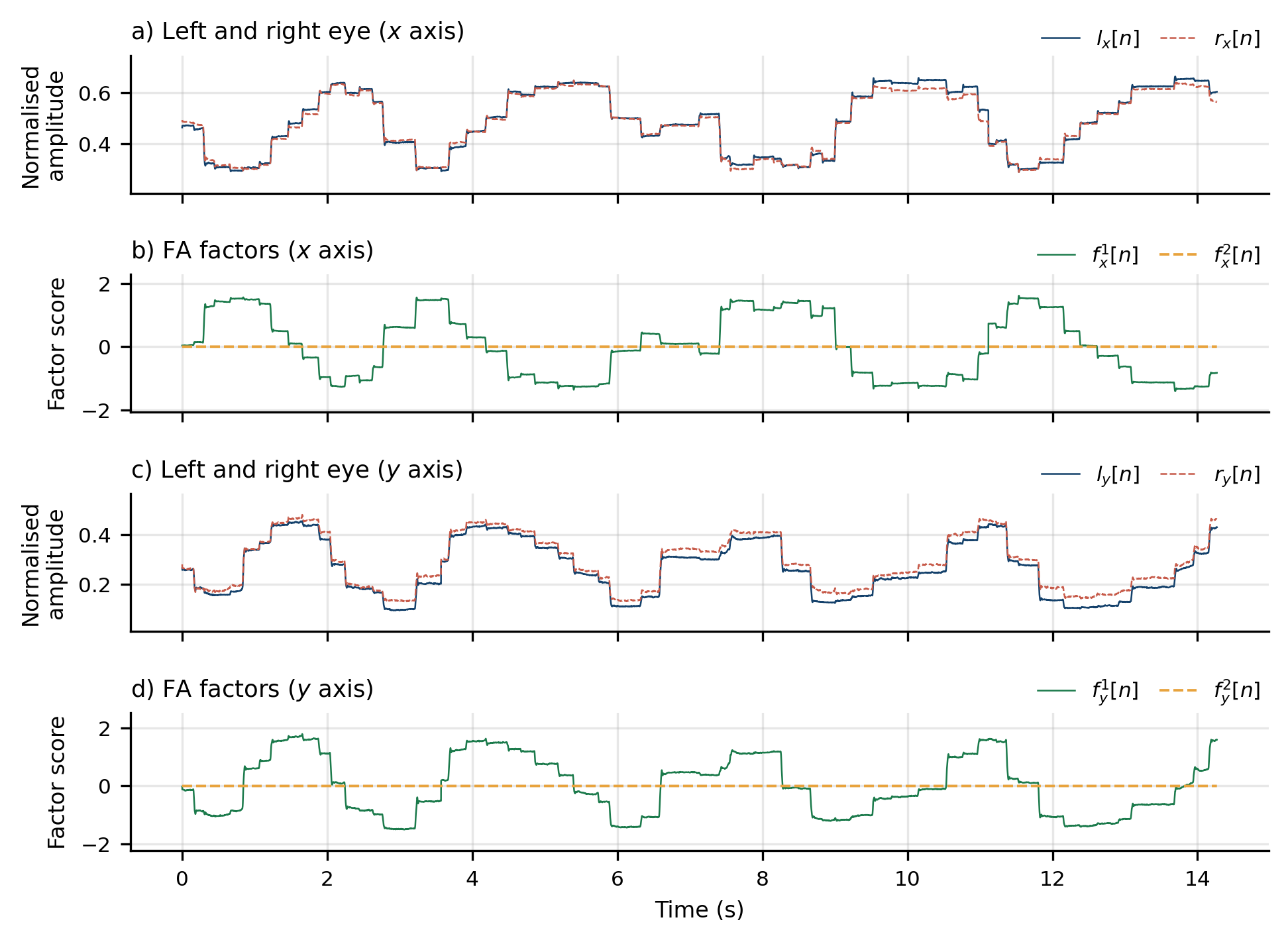}
\caption{Left and right eye channels $l_a[n]$, $r_a[n]$ (top of each pair) and the two \ac{FA} factors $f_a^1[n]$, $f_a^2[n]$ (bottom of each pair), for patient HF008, Expl.~1. Top pair: horizontal axis. Bottom pair: vertical axis. The first factor reproduces the shared scanpath; the second factor is flat at zero. The estimated noise variance $\psi$ of each channel is $0.40\%$ of its variance on the horizontal axis and $0.52\%$ on the vertical axis.}
\label{fig:fa_factors}
\end{figure}

\begin{figure}[htbp]
\centering
\includegraphics[width=\linewidth]{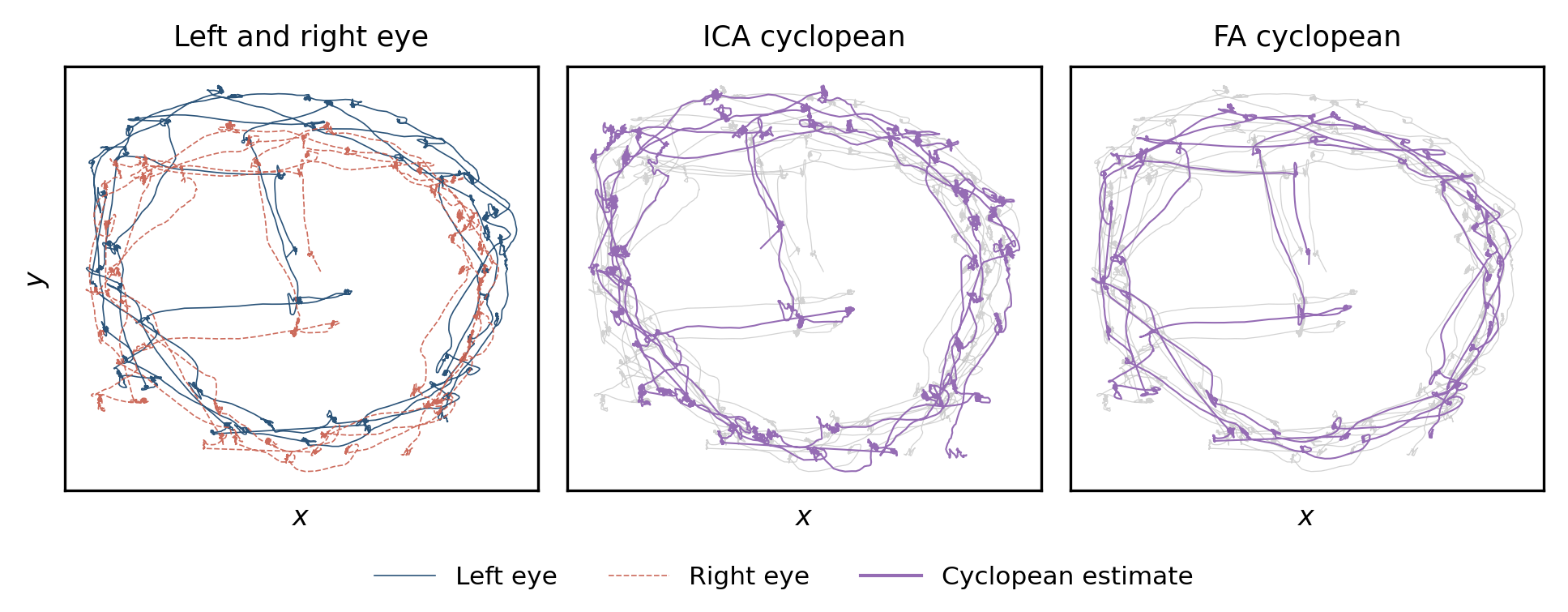}
\caption{Original left and right eye trajectories (left), \ac{ICA} cyclopean estimate (centre), and \ac{FA} cyclopean estimate (right), for patient HF008, Expl.~1. Both cyclopean estimates follow the shared circular scanpath, but differ from each other in detail.}
\label{fig:trajectory}
\end{figure}

\Cref{fig:fa_factors} shows, for patient HF008 during Expl.~1, the original left and right channels alongside the two \ac{FA} factors on both axes. The first factor $f_a^1[n]$ tracks the shared scanpath of the two eyes, while the second $f_a^2[n]$ remains flat at zero throughout, illustrating the corpus-wide null result above with a single recording. \Cref{fig:trajectory} contrasts the original two-eye trajectory with the \ac{ICA} and \ac{FA} cyclopean estimates for the same recording: both estimates follow the common scanpath, but differ from each other in detail.

\begin{figure}[htbp]
\centering
\includegraphics[width=\linewidth]{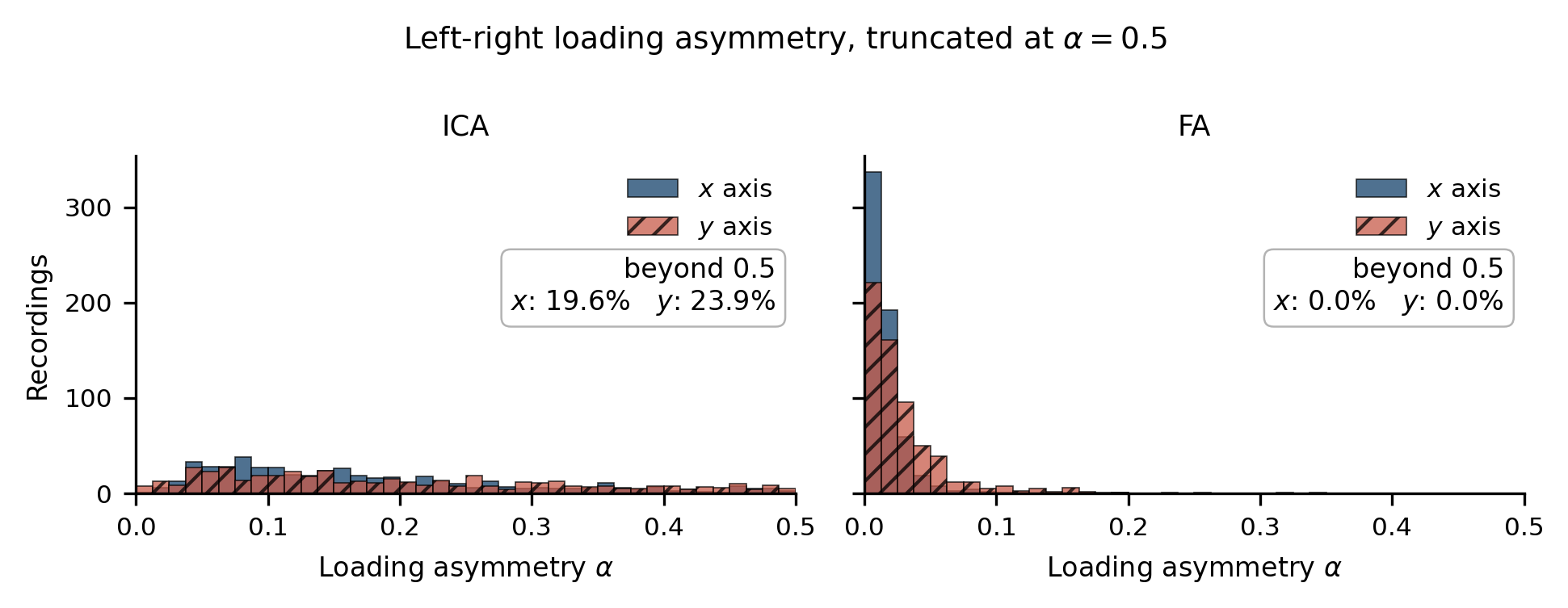}
\caption{Loading asymmetry $\alpha$ of \cref{eq:asym} for the selected \ac{ICA} component (left) and the first \ac{FA} factor (right), on both axes, aggregated over the corpus. Both panels share a vertical scale and are truncated at $\alpha = 0.5$, since the \ac{ICA} distribution extends to $\alpha$ close to $1$; the fraction of recordings falling beyond the cut is annotated in each panel. The \ac{ICA} distribution is broad and almost flat across the range shown, with a long tail toward strongly eye-dominated components; the \ac{FA} distribution concentrates sharply near zero and has no mass beyond the cut.}
\label{fig:ica_asymmetry}
\end{figure}

\Cref{fig:ica_asymmetry} compares the loading asymmetry $\alpha$ of the first \ac{FA} factor with that of the selected \ac{ICA} component (\cref{sec:res_ica}). The \ac{ICA} asymmetry is larger in the typical recording and far larger in the tail. For \ac{FA}, the median asymmetry is $0.012$ ($x$) and $0.019$ ($y$), and $0.041$ and $0.100$ at the 95th percentile. It never exceeds $\alpha = 0.5$, reaching at most $0.19$ ($x$) and $0.34$ ($y$).

Because \cref{eq:facyc} defines the cyclopean estimate directly from $w_l, w_r$, the near-zero loading asymmetry of the first factor is not a separate property of the signal but a restatement, at the level of the whole corpus, of how close $\hat{c}_a[n]$ stays to the \ac{VS}: close to a balanced binocular average in the typical recording, while remaining free to depart from it, on either axis and by as much as the data demand, in the recordings where the two eyes disagree.

\begin{figure}[htbp]
\centering
\includegraphics[width=\linewidth]{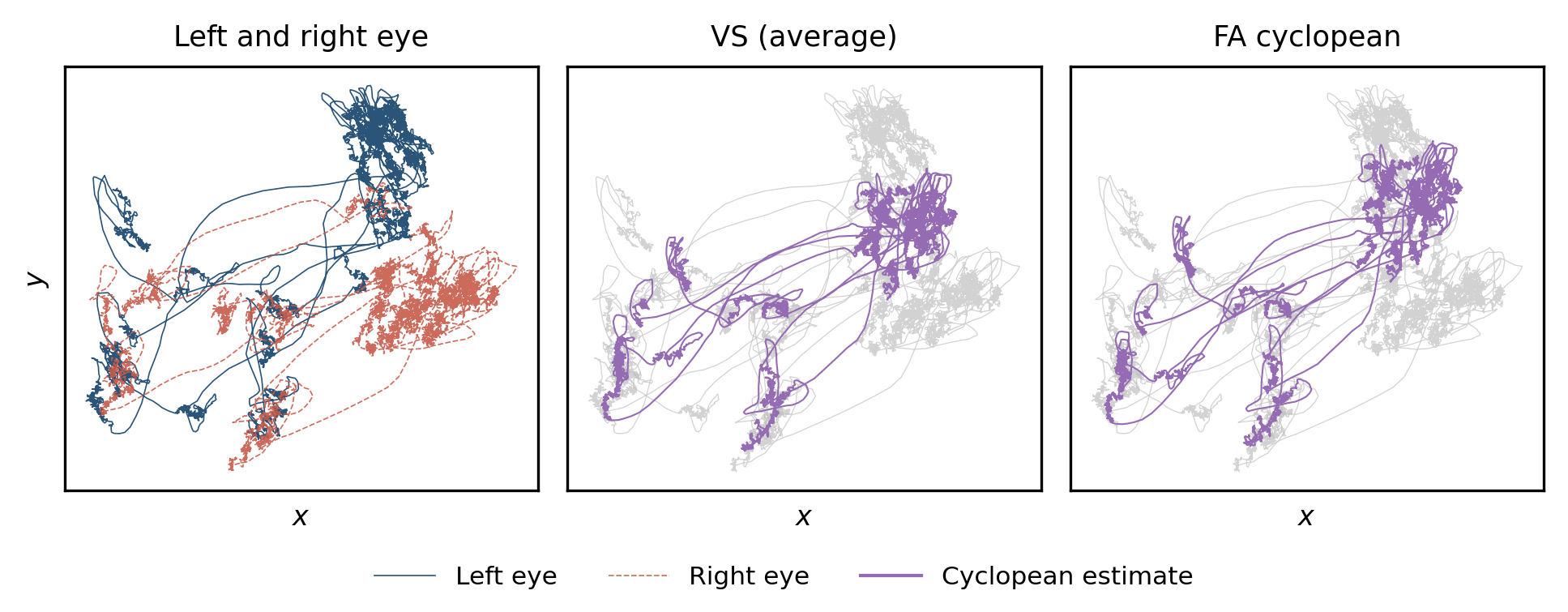}\\[0.5em]
\includegraphics[width=\linewidth]{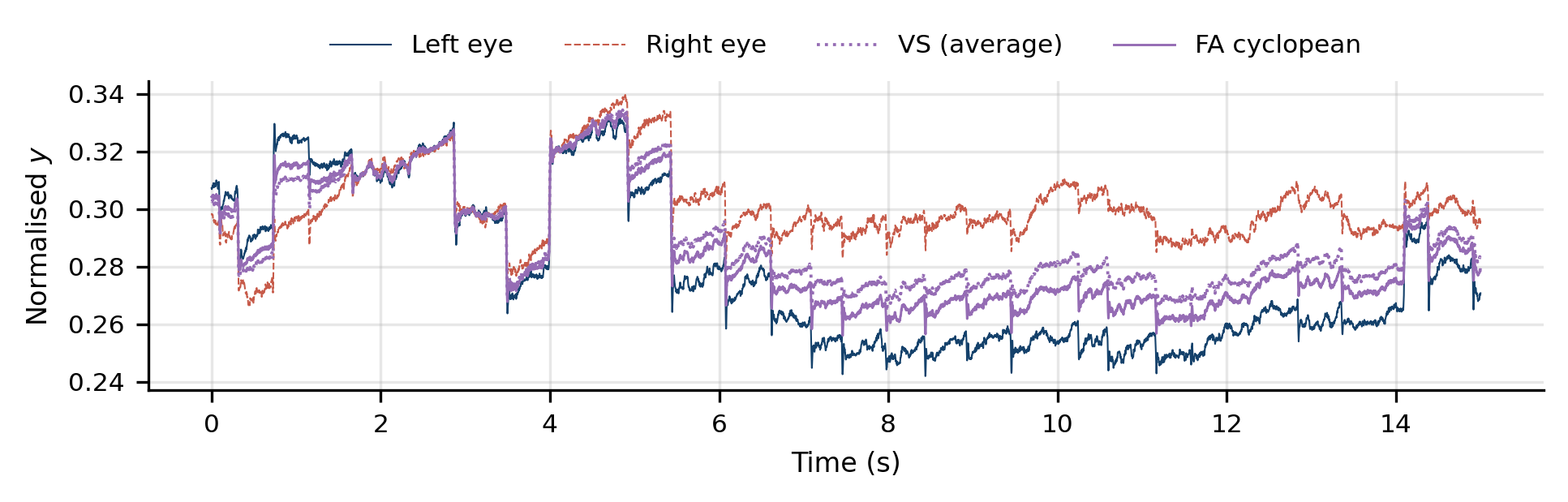}
\caption{The recording with the largest loading asymmetry in the corpus, HG070, Expl.~5. Top: original left and right eye trajectories (left), \ac{VS} (centre), and \ac{FA} cyclopean estimate (right). Bottom: vertical position over time for both eyes, the \ac{VS}, and the \ac{FA} estimate ($w_l = 0.67$, $w_r = 0.33$).}
\label{fig:extreme}
\end{figure}

\Cref{fig:extreme} shows the most extreme such recording in the corpus, HG070 during Expl.~5. On the horizontal axis the weights remain close to balanced ($\alpha_x = 0.05$). On the vertical axis the two eyes follow visibly different paths and the weights shift to $w_l = 0.67$ and $w_r = 0.33$ ($\alpha_y = 0.34$), so the \ac{FA} estimate lies closer to the left eye while the \ac{VS} stays halfway between the two. The recording passed every quality check of \cref{sec:preprocessing}, so the disagreement does not come from a failed measurement. $\alpha$ exceeds $0.1$ on at least one axis in 36 of the 628 recordings (5.7\%) and $0.2$ in 4 (0.6\%); these recordings show that combining the two eyes is more complex than it first appears, and that averaging is not the right answer by default.

\subsection{Event-level validation}
\label{sec:res_events}
\begin{figure}[htbp]
\centering
\includegraphics[width=0.6\linewidth]{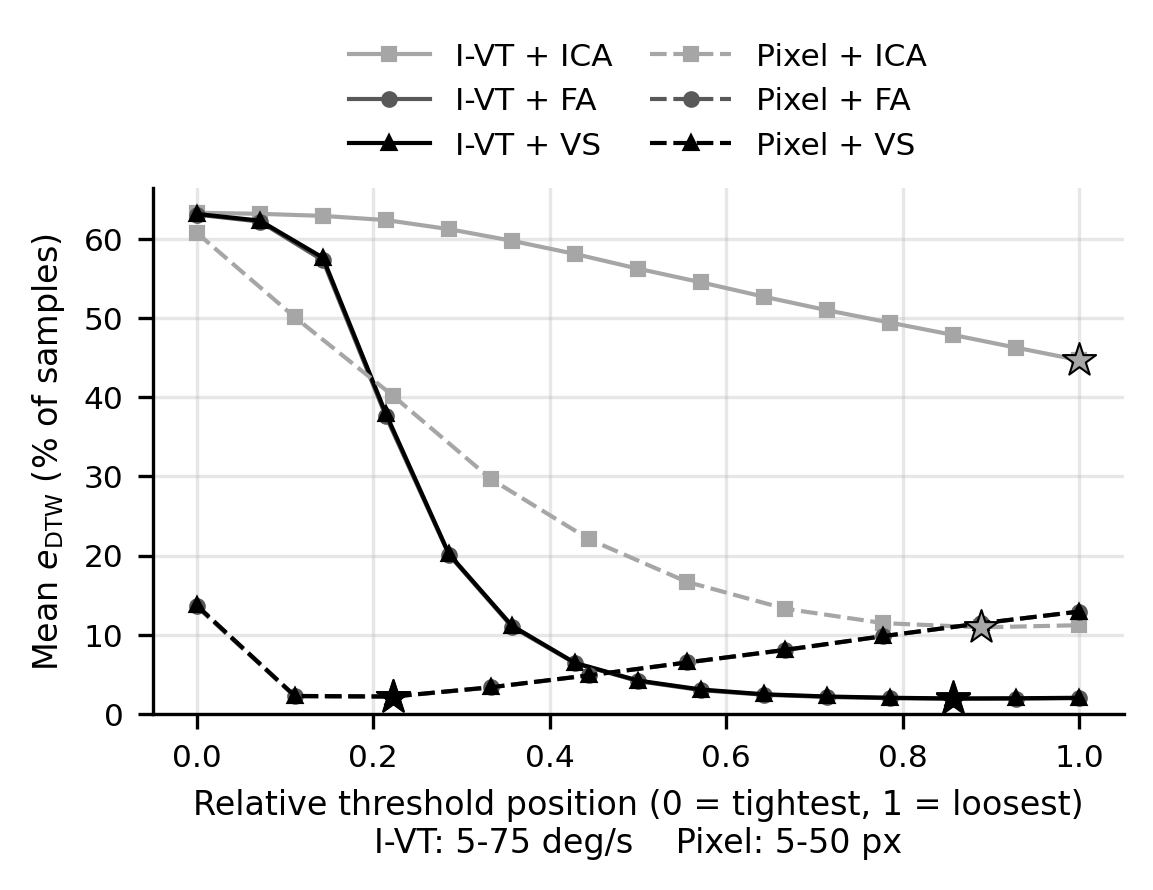}
\caption{Mean $e_{\mathrm{DTW}}$ over recordings (\cref{eq:dtwerr}) against threshold, for each of the detector-method combinations; the \ac{FA} and \ac{VS} curves almost coincide. The horizontal axis runs from the tightest to the loosest threshold tested for each detector, so the curves are comparable in relative terms; stars mark each combination's own minimum.}
\label{fig:threshold_all}
\end{figure}

For each detector, the threshold minimising the mean $e_{\mathrm{DTW}}$ of \cref{eq:dtwerr} was selected independently for \ac{ICA}, \ac{FA} and the \ac{VS}. Under \ac{IVT}, the optimum for \ac{FA} and for the \ac{VS} was $v_\theta = 65$\textdegree/s; the \ac{ICA} error was larger across the range swept and still decreasing at its end, so its nominal optimum, $75$\textdegree/s, is best read as the least unfavourable setting tested rather than a genuine minimum. Under pixel-dispersion detection, the optimum was $d=15$~px for \ac{FA} and the \ac{VS} and $d=45$~px for \ac{ICA}. \Cref{fig:threshold_all} places all six detector-method combinations on a common relative-threshold axis, from the tightest to the loosest value swept for each detector. The \ac{FA} and \ac{VS} curves almost coincide under both detectors; at their optima, \ac{IVT} \ac{FA} and \ac{VS} reach the lowest errors ($1.92\%$ and $1.96\%$), followed by pixel-dispersion \ac{FA} and \ac{VS} ($2.19\%$ and $2.20\%$), while \ac{ICA} remains higher under either detector ($10.94\%$ for pixel-dispersion and $44.68\%$ for \ac{IVT}), with \ac{IVT} \ac{ICA} above the other five throughout the range. \Cref{tab:events} reports the resulting event statistics, aggregated as mean~$\pm$~standard deviation across recordings at each method's own optimal threshold. The device parser, applied independently to the left and right channels, is the shared reference for both detectors.

\begin{table}[htbp]
\centering
\caption{Fixation and saccade statistics aggregated over recordings (mean~$\pm$~standard deviation), at each method's own best threshold. }
\label{tab:events}
\begin{tabular}{llcccc}
\toprule
Detector & Signal & \#fix & fix dur (ms) & \#sac & sac dur (ms) \\
\midrule
\multirow{5}{*}{\ac{IVT}}
 & Left eye (device)  & $42.0 \pm 9.3$ & $232.9 \pm 136.1$ & $38.2 \pm 9.8$ & $132.8 \pm 8.9$ \\
 & Right eye (device) & $41.8 \pm 9.4$ & $235.4 \pm 140.0$ & $38.0 \pm 9.8$ & $133.0 \pm 8.6$ \\
 & \ac{ICA} cyclopean & $14.3 \pm 18.8$ & $275.9 \pm 134.5$ & $127.9 \pm 123.6$ & $14.3 \pm 3.0$ \\
 & \ac{FA} cyclopean  & $40.1 \pm 9.9$ & $357.2 \pm 145.2$ & $52.1 \pm 19.6$ & $17.9 \pm 3.1$ \\
 & \ac{VS}            & $40.1 \pm 9.9$ & $356.0 \pm 144.8$ & $52.1 \pm 19.6$ & $17.9 \pm 3.1$ \\
\midrule
\multirow{5}{*}{Pixel}
 & Left eye (device)  & $42.0 \pm 9.3$ & $232.9 \pm 136.1$ & $38.2 \pm 9.8$ & $132.8 \pm 8.9$ \\
 & Right eye (device) & $41.8 \pm 9.4$ & $235.4 \pm 140.0$ & $38.0 \pm 9.8$ & $133.0 \pm 8.6$ \\
 & \ac{ICA} cyclopean & $36.4 \pm 9.1$ & $343.1 \pm 143.6$ & $35.4 \pm 9.1$ & $117.3 \pm 381.3$ \\
 & \ac{FA} cyclopean  & $43.0 \pm 8.2$ & $300.6 \pm 85.0$ & $42.0 \pm 8.2$ & $59.2 \pm 34.1$ \\
 & \ac{VS}            & $43.0 \pm 8.2$ & $299.8 \pm 85.2$ & $42.0 \pm 8.2$ & $59.3 \pm 33.8$ \\
\bottomrule
\end{tabular}
\end{table}

\begin{figure}[htbp]
\centering
\includegraphics[width=\linewidth]{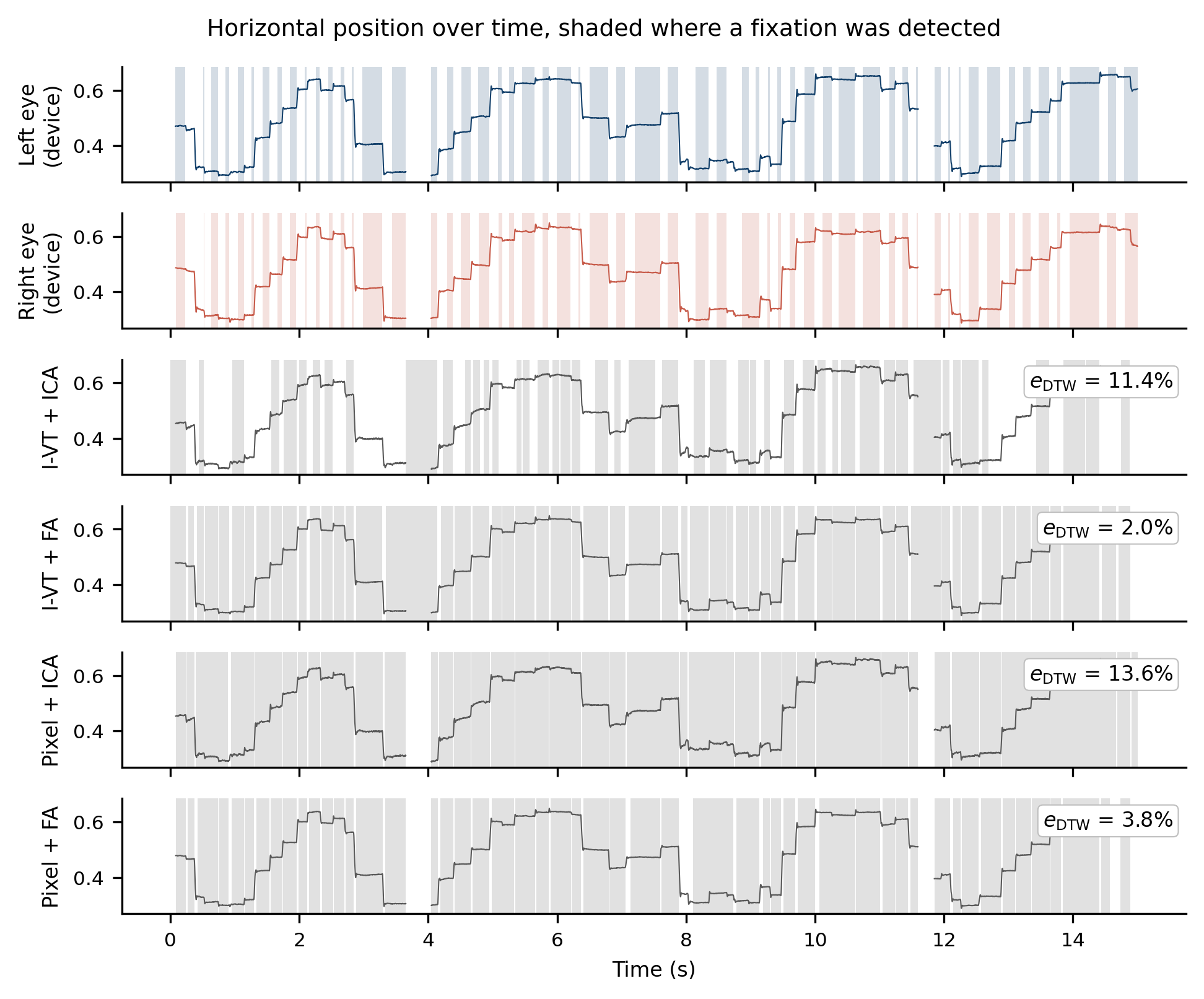}
\caption{Horizontal position over time for patient HF008, Expl.~1, with detected fixation intervals shaded, for the left eye, the right eye, and the \ac{ICA} and \ac{FA} cyclopean estimates under both detectors, each annotated with its $e_{\mathrm{DTW}}$ for this recording.}
\label{fig:fixations_scenarios}
\end{figure}

\Cref{fig:fixations_scenarios} shows the detected fixation intervals for patient HF008, Expl.~1, for the two device channels and for the \ac{ICA} and \ac{FA} estimates under both detectors, with each estimate's $e_{\mathrm{DTW}}$ for this recording: $2.0\%$ for \ac{FA} under \ac{IVT} and $3.8\%$ under pixel-dispersion detection, against $11.4\%$ and $13.6\%$ for \ac{ICA}.

Pooling individual fixation events across all recordings gives a clearer picture of duration than the per-recording mean. The device parser's median fixation duration is $155$~ms on both channels; the \ac{FA} estimate's median is longer under both detectors, $256$~ms under \ac{IVT} and $244$~ms under pixel-dispersion detection, the same as for the \ac{VS}, an increase of 57 to 65\%. The increase appears under both detectors. Since the reference events come from a different algorithm, it cannot be attributed to the cyclopean estimate alone.

The two methods also differ in how often event detection fails outright. At its own best threshold, \ac{ICA} produced zero detected fixations in $19.6\%$ of recordings under \ac{IVT}, and $2$ or fewer in $44.6\%$, against $0.0\%$ and $0.2\%$ for \ac{FA} and the \ac{VS}; under pixel-dispersion detection the gap nearly closes, at $0.3\%$ and $0.5\%$ for \ac{ICA} against $0.0\%$ and $0.0\%$ for \ac{FA} and the \ac{VS}, which matches the device parser's own baseline of no empty recordings. This is a robustness argument distinct from, though related to, the identifiability results of \cref{sec:res_ica}: even at a threshold chosen specifically to favour it, \ac{ICA} fails outright far more often than \ac{FA} does under \ac{IVT}. \Cref{tab:fixcounts} gives the full per-recording fixation-count distribution underlying this comparison, binned more coarsely (0, 1--4, 5--10, 11+ fixations) than the individual thresholds quoted above.

\begin{table}[htbp]
\centering
\caption{Distribution of per-recording fixation counts, at each method's own best threshold (as in \cref{tab:events}).}
\label{tab:fixcounts}
\begin{tabular}{lccccc}
\toprule
Signal & $n$ & 0 fix (\%) & 1--4 fix (\%) & 5--10 fix (\%) & 11+ fix (\%) \\
\midrule
Left eye (device)   & 628 & 0.0  & 0.0  & 0.3  & 99.7 \\
Right eye (device)  & 628 & 0.0  & 0.0  & 0.3  & 99.7 \\
\ac{IVT} + \ac{ICA}  & 628 & 19.6 & 35.5 & 11.0 & 33.9 \\
\ac{IVT} + \ac{FA}   & 628 & 0.0 & 0.2 & 0.5 & 99.4 \\
\ac{IVT} + \ac{VS}   & 628 & 0.0 & 0.2 & 0.5 & 99.4 \\
Pixel + \ac{ICA}     & 628 & 0.3 & 0.5 & 0.6 & 98.6 \\
Pixel + \ac{FA}      & 628 & 0.0 & 0.0 & 0.0 & 100.0 \\
Pixel + \ac{VS}      & 628 & 0.0 & 0.0 & 0.0 & 100.0 \\
\bottomrule
\end{tabular}
\end{table}

\begin{figure}[htbp]
\centering
\includegraphics[width=0.6\linewidth]{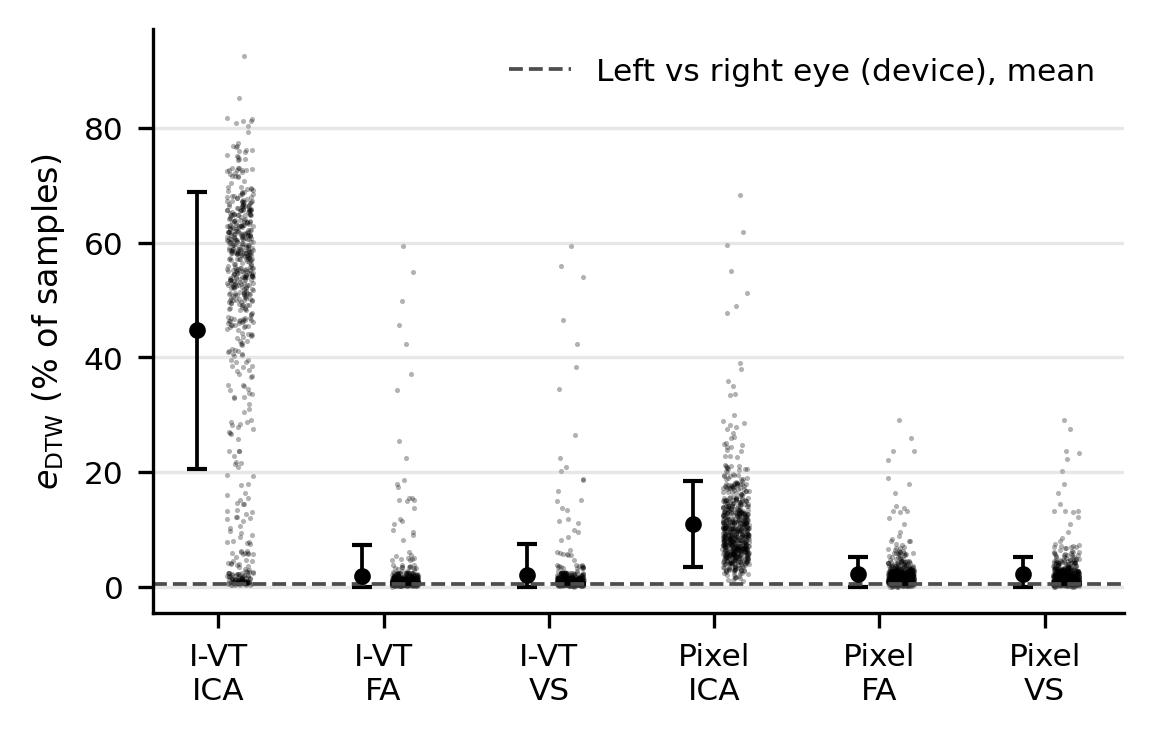}
\caption{$e_{\mathrm{DTW}}$ of \cref{eq:dtwerr} at each method's best threshold: $\mu \pm \sigma$ over the 628 recordings (bars truncated at 0), and each recording as a dot; the dashed line marks the mean inter-eye baseline $e_{\mathrm{DTW}}^{lr}$.}
\label{fig:event_errors}
\end{figure}

\Cref{fig:event_errors} gives the distribution of $e_{\mathrm{DTW}}$ over recordings at each method's selected threshold. The inter-eye baseline $e_{\mathrm{DTW}}^{lr}$ is $0.54 \pm 1.11\%$ ($\mu \pm \sigma$), marked by the dashed line.

\section{Discussion}
\label{sec:discussion}

The results show that \ac{ICA} is not well suited to estimating the cyclopean component during free visual exploration. In some recordings, neither source correlated strongly with the original eye signals, reflecting cases where the decomposition did not yield a clearly identifiable cyclopean component. \ac{ICA} also showed substantially higher \ac{DTW} errors than FA and the \ac{VS}, under both IVT and pixel-dispersion detection.

The FA estimate remains close to the \ac{VS} while allowing unequal eye weights when the two channels disagree. Although the weights are nearly balanced in most recordings, their variability highlights the flexibility of \ac{FA} and suggests that equal weighting may not be appropriate in every case. The consistently null second factor further suggests that a single shared factor is sufficient, with inter-eye discrepancies captured by the unique-noise terms.

At the event level, \ac{FA} and the \ac{VS} achieve comparable \ac{DTW} errors, substantially below those of \ac{ICA}. However, both produce longer fixation durations than the device parser, and the optimal thresholds differ across methods. Thresholds therefore cannot be assumed to transfer directly between cyclopean estimates and the device parser.

A previous study found that \ac{ICA} recovers the cyclopean component reliably under smooth pursuit~\cite{luque2024cyclopean}. Under free exploration it often does not, while \ac{FA} does in every recording. The two tasks produce different signals: under smooth pursuit, removing the target trajectory leaves a residual dominated by large, non-Gaussian saccadic events, which \ac{ICA} needs to separate its sources. Free exploration has no target trajectory to remove, and its recorded channels are close to Gaussian (\cref{fig:gaussianity_check}). 

Finally, evaluation is necessarily indirect because the cyclopean eye is a virtual signal with no measurable ground truth. Comparing detected fixations with those of the device parser provides a reference for evaluating the estimates, but cannot establish the accuracy of the estimated cyclopean component itself. Within this limitation, FA provides a robust approach to estimating the shared binocular component during free visual exploration.

\section{Conclusions}

This study evaluated the feasibility of estimating the cyclopean eye signal during binocular free visual exploration, a target-free paradigm relevant to the screening and monitoring of PD. While \ac{ICA} previously succeeded in target-driven tasks like smooth pursuit, the absence of a target trajectory under free exploration severely restricts its applicability, leading to ambiguous component selection in up to $30.2\%$ of recordings and substantial degradation in fixation and saccade structure. 

In contrast, FA overcomes these limitations by isolating the shared binocular variance without requiring non-Gaussian signal assumptions. Across 628 visual explorations from 112 participants, FA provided a fully identifiable cyclopean estimate in all cases. The empirical finding of a consistently null second factor confirms that a single shared latent factor is sufficient to account for common gaze dynamics, absorbing eye-specific discrepancies into independent noise terms. At the event level, the FA estimate preserved oculomotor structure with low \ac{DTW} alignment errors ($1.92\%$ under I-VT and $2.19\%$ under pixel-dispersion detection), matching the performance of the traditional \ac{VS} while offering the critical advantage of dynamically adjusting eye weights during binocular asymmetry. 

Overall, these findings evidence that \ac{FA} provides a robust, fully automatic, and interpretable framework for recovering a unified gaze signal in unconstrained visual exploration tasks, paving the way for more accurate oculomotor biomarker extraction in clinical populations.

\section*{Acknowledgments}
Work supported by the Ministry of Economy and Competitiveness of Spain (PID2021-128469OB-I00), by Ministry of Health of Spain (FIS PI21/00771), and by Comunidad de Madrid, Spain. This is also supported by the MISTI MIT Global Experiences. Universidad Politécnica de Madrid also supported J. D. Arias-Londoño through a María Zambrano grant (UP2021-035) funded by the European Union-NextGenerationEU.
The authors would also like to thank the Madrid ELLIS unit (European Laboratory for Learning \& Intelligent Systems) for its indirect support.
Finally, the authors also acknowledge Universidad Politécnica de Madrid for the computing resources provided on the Magerit Supercomputer.

\subsection*{Ethics Declaration}
The study was approved by the Ethics Review Boards of the HUF and HGUGM, with codes 18/11-ENM1 and 11/2015 respectively, in accordance with the Spanish Ethical Review Act. All participants provided written informed consent.

\section*{Declaration of generative AI and AI-assisted technologies in the writing process}

During the preparation of this manuscript, the authors used commercially available LLMs to improve the English language and enhance the clarity and style of the text. Following the use of this tool, the authors carefully reviewed and edited the content as necessary and assume full responsibility for the accuracy, integrity, and content of the published work.

\section*{Declaration of competing interest}
The authors declare that they have no competing financial interests or personal relationships that could be perceived as having influenced the research, authorship, or publication of this article.

\bibliographystyle{unsrt}
\bibliography{references}

\end{document}